\documentclass[12pt]{article}
\usepackage{epsfig}
\usepackage{amssymb}
\usepackage{amsmath}
\usepackage{bm}                  
\usepackage{cancel}
\usepackage{cite}
\usepackage{comment} 
\newcommand{\pres}[2]{\setbox0=\hbox{$\scriptstyle #1$} \dimen0=\dp0  
              \dimen1=\ht0 \divide\dimen1 by 3
              \advance\dimen0 by \dimen1
              \hbox{\lower\dimen0 \hbox{$\scriptstyle #2\!\!$}} #1}
\newcommand{\nc}{\newcommand}
\nc{\figcap}[1]{\begin{quote}\refstepcounter{figure}
        {\bf Figure \thefigure}: {\small #1}\end{quote}}
\nc{\fig}[1]{\mbox{Fig.~\ref{#1}}}
\nc{\noi}{\noindent}
\nc{\bea}{\begin{eqnarray}}
\nc{\eea}{\end{eqnarray}}
\nc{\bean}{\begin{eqnarray*}}
\nc{\eean}{\end{eqnarray*}}
\nc{\ba}{\begin{array}}
\nc{\ea}{\end{array}}
\nc{\be}{\begin{equation}}
\nc{\ee}{\end{equation}}
\nc{\nn}{\nonumber}
\nc{\bra}[1]{\langle #1|}
\nc{\ket}[1]{|#1\rangle}
\nc{\av}[1] {\langle #1\rangle}
\nc{\vac}[1] {\langle 0| #1|0\rangle}
\nc{\amp}[2]{\langle #1|#2\rangle}
\nc{\da}{\dagger}
\nc{\pa}{\partial}
\nc{\ga}{\gamma}
\nc{\ep}{\epsilon}
\nc{\tf}{t_f}
\nc{\half}{\ensuremath{\frac{1}{2}}}
\nc{\hHH}{\hat H}
\nc{\ha}{\hat a}
\nc{\hO}{\hat O}
\nc{\hAA}{\hat A}
\nc{\hB}{\hat B}
\nc{\hG}{\hat G}
\nc{\hN}{\hat N}
\nc{\hU}{\hat U}
\nc{\hx}{\hat{x}}
\nc{\hp}{\hat{p}}
\nc{\hpsi}{\hat \psi}
\nc{\hphi}{\hat \phi}
\nc{\hpi}{\hat \pi}
\nc{\hpd}{\hat \psi ^\dagger}
\nc{\hE}{\hat E}
\nc{\hb}{\hat b}
\nc{\hc}{\hat c}
\nc{\hjo}{\hat j _0}
\nc{\hrho}{\hat \rho}
\nc{\leave}{\! \! \! \! \! / \, \,}
\nc{\intl}[1]{\int d\! #1 \,} 
\nc{\intll}[3]{\int _#1^#2 d\! #3 \,} 
\nc{\lm}{\lim _{y \rightarrow x}}
\nc{\scd}{\partial ^2 _{A_T}}
\nc{\fd}[1]{\frac{\delta }{\delta #1}} 
\nc{\pad}[1]{\frac{\partial}{\partial #1}} 
\nc{\refpa}[1]{(\ref{#1})} 
\nc{\calH}{\ensuremath{\mathcal{H}}}
\nc{\calD}{\ensuremath{\mathcal{D}}}
\nc{\calL}{\ensuremath{\mathcal{L}}}
\nc{\calO}{\ensuremath{\mathcal{O}}}
\nc{\hcalO}{\ensuremath{\hat \mathcal{O}}}
\nc{\calK}{\ensuremath{\mathcal{K}}}
\nc{\Tr}{\ensuremath{\mathrm{Tr}}}
\nc{\tr}{\ensuremath{\mathrm{tr}}}
\nc{\ra}{\rightarrow}
\nc{\lr}{\leftrightarrow}
\nc{\phistar}{\phi^*}
\nc{\etat}{\eta_T}
\nc{\het}{\hat E_T}
\nc{\hpt}{\hat \psi_T}
\nc{\hpdt}{\hat \psi ^\dagger_T}
\nc{\bart}{\bar{t}}
\nc{\barp}{\bar{p}}
\nc{\barT}{\bar{T}}
\nc{\hbarrho}{\hat{\bar{\rho}}}
\nc{\bga}{\ensuremath{\mbox{\boldmath{$\gamma$}}}}
\nc{\bsi}{\ensuremath{\mathbf{\sigma}}}
\nc{\bx}{\ensuremath{\mathbf{x}}}
\nc{\by}{\ensuremath{\mathbf{y}}}
\nc{\bz}{\ensuremath{\mathbf{z}}}
\nc{\bp}{\ensuremath{\mathbf{p}}}
\nc{\bn}{\ensuremath{\mathbf{n}}}
\nc{\bbp}{\ensuremath{\bar{\mathbf{p}}}}
\nc{\bP}{\ensuremath{\mathbf{P}}}
\nc{\hbA}{\hat{\ensuremath{\mathbf{A}}}}
\nc{\hbB}{\hat{\ensuremath{\mathbf{B}}}}
\nc{\bA}{\ensuremath{\mathbf{A}}}
\nc{\bJ}{\ensuremath{\mathbf{J}}}
\nc{\bB}{\ensuremath{\mathbf{B}}}
\nc{\bH}{\ensuremath{\mathbf{H}}}
\nc{\bM}{\ensuremath{\mathbf{M}}}
\nc{\bD}{\ensuremath{\mathbf{D}}}
\nc{\bE}{\ensuremath{\mathbf{E}}}
\nc{\hbE}{\hat{\ensuremath{\mathbf{E}}}}
\nc{\br}{\ensuremath{\mathbf{r}}}
\nc{\bj}{\ensuremath{\mathbf{j}}}
\nc{\bOm}{\ensuremath{\mathbf{\Om}}}
\nc{\om}{\omega}
\nc{\Om}{\Omega}
\nc{\sgn}{\mbox{sgn}}
\nc{\deltabar}{\mbox{$\delta\hspace*{-8pt}\vspace*{-8pt}-$}}
\nc{\gammat}{\tilde{\gamma}}
\nc{\mub}{\bar{\mu}}
\nc{\rhob}{\bar{\rho}}
\nc{\Bb}{\bar{B}}
\nc{\Jb}{\bar{J}}
\nc{\Mb}{\bar{M}}
\nc{\Tb}{\bar{T}}
\nc{\sbar}{\bar{s}}
\nc{\betab}{\bar{\beta}}
\nc{\hj}{\hat j}
\nc{\hQ}{\hat Q}
\nc{\hJ}{\hat J}
\nc{\hA}{\hat A}
\nc{\hH}{\hat H}
\nc{\de}{\delta}
\nc{\leri}{\leftrightarrow}
\nc{\llabel}[1]{\label{#1}\marginpar{#1}}
\nc{\bc}{\begin{center}}
\nc{\ec}{\end{center}}
\nc{\inv}[1]{\frac{1}{#1}}

\newlength{\overeqskip}
\newlength{\undereqskip}
\nc{\eq}[1]{\mbox{Eq.~(\ref{#1})}}
\nc{\eps}{\epsilon}
\nc{\goto}{\rightarrow}
\nc{\cF}{{\cal F}}
\nc{\cG}{{\cal G}}
\nc{\cH}{{\cal H}}
\newcounter{sectionc}
\newcounter{subsectionc}
\newcounter{subsubsectionc}
\renewcommand{\section}[1]
{\refstepcounter{sectionc}\vspace{0.0cm}
\setcounter{subsectionc}{0}\setcounter{subsubsectionc}{0}\noindent 
	{\bf\thesectionc. #1}}
\renewcommand{\subsection}[1] {\vspace{0.0cm}
\addtocounter{subsectionc}{1} 
	\setcounter{subsubsectionc}{0}\noindent 
	{\it\thesectionc.\thesubsectionc. #1}\par\vspace{0.4cm}}
\renewcommand{\subsubsection}[1] {\vspace{0.6cm}\addtocounter{subsubsectionc}{1}
	\noindent {\rm\thesectionc.\thesubsectionc.\thesubsubsectionc. 
	#1}\par\vspace{0.0cm}}

\renewcommand{\theequation}{\thesectionc.\arabic{equation}}
\renewenvironment{thebibliography}[1]
	{\begin{list}{\arabic{enumi}.}
	{\usecounter{enumi}\setlength{\parsep}{0pt}
\setlength{\leftmargin 1.25cm}{\rightmargin 0pt}
	 \setlength{\itemsep}{6pt} \settowidth
	{\labelwidth}{#1.}\sloppy}}{\end{list}}
\newcommand{\seqnoll}{\setcounter{equation}{0}}
\begin{document}
%
%
%
%
%
\setlength{\jot}{10pt} 
%
%
\thispagestyle{empty} 
%
%
%
\vspace{1cm}
\begin{center}  
\baselineskip 1.2cm 
{\Huge\bf  Quantum Field Theory with Classical Sources - Linearized Quantum Gravity
}
\\[5mm]  
\normalsize 
\end{center} 
{\centering 
{\large Bo-Sture K. Skagerstam\footnote{Corresponding author. Email address: bo-sture.skagerstam@ntnu.no}$^{,a)}$}, 
{\large Karl-Erik Eriksson\footnote{Email address: frtkee@chalmers.se}$^{,b)}$}, 
{\large Per K. Rekdal\footnote{Email address: per.k.rekdal@himolde.no}$^{,c)}$} 
\\[5mm] 
$^{a)}$Department of Physics, NTNU,  Norwegian University of Science and Technology, N-7491 Trondheim, Norway
\\[1mm]
$^{b)}$ Department of Space, Earth and Environment,
Chalmers University of Technology, SE-412 96 G\"{o}teborg, Sweden
\\[1mm]
$^{c)}$Molde University College, P.O. Box 2110,  N-6402 Molde, Norway
\\[1mm]
} 
%
%
%
%
%
%
%
%
%
%
%
\begin{abstract} 
%
%
\indent  In a previous work \cite{SER_QED} and in terms of an exact quantum-mechanical framework,   $\hbar$-independent causal and retarded expectation values of the second-quantized electro-magnetic fields in the Coulomb gauge were derived   in the presence of a conserved classical electric current.    The classical $\hbar$-independent Maxwell's equations then naturally emerged.  In the present work,  we extend these considerations  to linear gravitational quantum deviations around a flat Minkowski space-time in a Coulomb-like gauge. The emergence of the classical causal and properly retarded linearized classical theory of general relativity with a conserved classical energy-momentum tensor is then outlined. The quantum-mechanical framework  also provides for a simple approach to classical quadrupole gravitational radiation of Einstein and   microscopic spontaneous graviton emission and/or absorption processes.   
\vspace{1mm}
\end{abstract} 
%
%
%
%
%
%
\vspace{0.5cm}
\newpage
\setcounter{page}{1}
\seqnoll
\bc{
\section{\large Introduction}
\label{sec:introuction}
}\ec
%
%

In electro-dynamics  it is natural to introduce gauge-dependent scalar and vector potentials. These potentials do not have to be local  in space and  time. 
It can then be a rather delicate issue to verify that  gauge-independent observables obey the physical constraint of causality and that they also are properly retarded. Attention to this issue is often discussed in a classical framework. For instance one  then  shows in what manner  various choices of gauge  give rise to the same electro-magnetic field strengths (for recent discussions see, e.g.,  Refs.\cite{brill_67,gardiner_88,Jackson_75,rohrlich_01,jackson_02,stewart_03,hnizdo_04,heras_07,millington_2016}).

In a previous publication \cite{SER_QED} we have shown that the time-evolution as dictated by quantum mechanics, for second-quantized electro-magnetic fields in the presence of a classical conserved electric current,  automatically solves these issues. The classical theory of Maxwell then naturally emerges in line with more general $S$-matrix arguments due to Weinberg \cite{weinberg_1965}. We  also showed that various exact $\hbar$-independent radiative processes like the classical  Vavilov-\v{C}herenkov radiation \cite{Cherenkov_34, {Frank_Tamm_37}} can be obtained in a straightforward manner \cite{Ginzburg_1940, Harris_72, Marcuse_80}.

In the present paper we extend the work of Ref.\cite{SER_QED} to  linear gravitational quantum deviations around a flat Minkowski space-time in a Coulomb-like gauge in the presence of a classical and conserved energy-momentum tensor. The time-evolution as dictated by quantum mechanics, for a suitably defined second-quantized  gravitational field,  then automatically solves the issues of causality and retardation in a similar manner as in the case of electro-dynamics. It will, however, be argued that this is only true for  certain propagating physical degrees of freedom and not for all degrees of freedom of the gravitational field as is sometimes claimed in the literature (see, e.g., Ref.\cite{rohrlich_01}, and Ref.\cite{rothman_2018} for various historical remarks). The corresponding Coulomb-like degrees of freedom, like the Newton's gravitational potential, are then neither retarded nor causal. In fact, they are shown to be instantaneous.
The   dynamical equations can, as in the case of electro-dynamics \cite{SER_QED}, be reduced to a system of decoupled harmonic oscillators with space-time dependent external forces. 
No pre-defined global causal order is assumed other than the deterministic time-evolution as prescribed by the Schr\"{o}dinger equation. The classical theory of Einstein then naturally emerges  in terms of  expectation values of  the second-quantized gravitational field for any initial quantum state again in line with more general $S$-matrix arguments due to Weinberg \cite{weinberg_1965} and by Boulware and Deser \cite{deser_75}.

   The paper  is organized as follows. In Section  \ref{sec:Einstein} we recall, for reasons of completeness,  the classical version of Einstein's theory of weak gravitational fields in vacuum   in the presence of a space-time dependent conserved energy-momentum tensor, and a proper set of propagating degrees of freedom is constructed. 
 The quantized degrees of freedom  around a flat Minkowski space-time in the presence of a conserved  space-time dependent energy-momentum tensor is outlined in Section \ref{sec:multimode}. In this section we also make some comments on the construction of a conserved energy-momentum tensor for  these quantized gravitational degrees of freedom and the Weinberg-Witten theorem \cite{WW_1980}. In Section \ref{sec:causality} we consider the issues of causality and retardation in terms of quantum-mechanical averages of the second-quantized gravitational degrees of freedom and geodesic deviation. The emergence of the classical weak-field limit of Einstein's general theory of relativity is then outlined. 
In Section \ref{sec:grav_emission_proc}   we also discuss, in a quantum-mechanical framework, Einstein's classical  gravitational quadrupole radiation as well as the microscopic spontaneous graviton emission process from a quantum source in terms of an excited hydrogen-like atom. In Section \ref{sec:final_remarks} finally,  we present some conclusions and remarks.  In an Appendix, we present some calculational techniques as made use of  in obtaining decay rates for the emission of gravitons.
\newline

%
%
\vspace{0.5cm}
%
\bc{
\section{\large Weak Classical Gravitational Fields}
\label{sec:Einstein}
}\ec
%
%

We  consider classical weak deviations from the flat Minkowski space-time in terms of the metric tensor, i.e., we write
\begin{equation} \label{grav1}
    g_{\mu \nu} = \eta_{\mu \nu} + h_{\mu \nu} ~ ,
\end{equation}
where $|h_{\mu\nu}| \ll 1$ for all $\mu$ and $\nu$. We will make use of the conventions that Greek indices run from $1$ to $4$, with the space-time 
$x$-coordinate components $x^{\mu} = (x^1,x^2,x^3,x^4) \equiv ({\bf x},x^4=ct)$ as well as $\partial_{\mu}=\partial/\partial x^{\mu} $, and the diagonal metric $\eta_{\mu \nu}$ has the signature $(1,1,1,-1)$. 
%
%
%
The Minkowski metric $\eta_{\mu \nu}$ is used to raise or lower space-time Greek indices. When convenient, we will make use of the notation $f(x)\equiv f({\bf x},t)$ for space-time dependent fields. Apart from using the index $4$ instead of $0$ for time components,  we follow the conventions of Ref.\cite{Weinberg_1972}.

In general, the fundamental classical Einstein field equation takes the form
\begin{equation} \label{grav2}
     G_{\mu \nu} \equiv R_{\mu \nu} - \frac{1}{2}\eta_{\mu \nu}R = - \kappa T_{\mu \nu} ~ ,
\end{equation}
with 
\begin{equation} \label{grav3}
     \kappa = 8 \pi G/c^4 ~ .
\end{equation}
In the weak-field limit,  the Riemann-Christoffel curvature tensor $R_{\lambda\mu\nu\kappa}$  is approximated by
\begin{equation} \label{Riemann_Chrisfoffel}
    R^{(1)}_{\lambda\mu\nu\kappa} =\frac{1}{2}\left ( 
\frac{\partial^2 h_{\lambda\nu}}{\partial x^{\kappa}\partial x^{\mu}} 
- \frac{\partial^2 h_{\mu\nu}}{\partial x^{\kappa}\partial x^{\lambda}}
- \frac{\partial^2 h_{\lambda\kappa}}{\partial x^{\mu}\partial x^{\nu}}
+ \frac{\partial^2 h_{\mu\kappa}}{\partial x^{\nu}\partial x^{\lambda}}
\right)~ .
\end{equation}
In general,  $R_{\mu \nu}\equiv   R^{\lambda}_{\,\,\, \mu\nu\lambda}$ as well as $R\equiv R^{\lambda}_{\,\,\,\lambda} $. The Einstein tensor $G_{\mu \nu}$ is therefore approximated by $G_{\mu\nu}^{(1)}$ as defined by

%
%
\begin{equation} \label{grav4}
    G_{\mu\nu}^{(1)} = \frac{1}{2}\bigg ( \delta_{\mu}^{\alpha} \delta_{\nu}^{\beta} \square + \eta^{\alpha \beta} \partial_{\mu} \partial_{\nu} - \delta_{\nu}^{\beta} \partial_{\mu} \partial^{\alpha} - \delta_{\mu}^{\alpha} \partial_{\nu} \partial^{\beta} + \eta_{\mu \nu} \Big ( \partial^{\alpha} \partial^{\beta} - \eta^{\alpha \beta} \square \Big ) \Bigg ) h_{\alpha \beta} ~ ,
\end{equation}
with $\square \equiv \partial^{\alpha} \partial_{\alpha} = \nabla^2 - \partial^2/\partial (ct)^2$.
 $G_{\mu\nu}^{(1)}$ is then such that  $\partial^{\mu}G_{\mu\nu}^{(1)} =0$ and therefore we must  impose conservation of the energy-momentum tensor, i.e.,   $\partial^{\mu}T_{\mu \nu} =0$.  

The Riemann-Christoffel curvature tensor Eq.(\ref{Riemann_Chrisfoffel})   has  a local gauge invariance, i.e., it is invariant under infinitesimal coordinate transformations 
\begin{equation}  \label{grav5}
     x_{\mu} \rightarrow x'_{\mu} = x_{\mu} - \xi_{\mu} ~ ,
\end{equation}
where we  only consider $\xi_{\mu}$ and terms like $\partial_{\nu} \xi_{\mu}$ to first order. 
This is so since Eq.(\ref{grav5}) and the tensor properties of $g_{\mu \nu}$  imply 
\begin{equation} \label{grav6}
    h_{\mu \nu} \rightarrow h'_{\mu \nu} = h_{\mu \nu} + \partial_{\mu} \xi_{\nu} + \partial_{\nu} \xi_{\mu} ~ .
\end{equation}
It then follows that  $G_{\mu\nu}^{(1)}$ has the same invariance.
For our purposes, we impose Lorentz-covariant equations of motion and regard (\ref{grav6}) as a conventional local gauge-transformation of  the gauge-field  degrees of freedom $h_{\mu \nu}$. 
\newline
\indent It is now well-known that the components $G_{\mu 4}^{(1)} = - \kappa T_{\mu 4}$ of Einstein's equation Eq.(\ref{grav2}) in the weak-field limit, required to have conservation of the energy-momentum tensor, i.e., $\partial^{\mu}T_{\mu\nu} = 0$,  contain at most time-derivatives of first order and can therefore be used as initial conditions for second-order equations of motion  (also see, e.g., Section 7 in Ref.\cite{Weinberg_1972} and for  more recent discussions Refs.\cite{Bertschinger_1996,Flanagan_2005}).  We therefore consider the components $G_{\mu 4}^{(1)} = - \kappa T_{\mu 4}$ to be constraint equations  similar to the constraint equation $\nabla \cdot {\bf E} = - \nabla^2 \phi = \rho/\epsilon_0$ in electro-dynamics in the Coulomb gauge (see, e.g., Ref.\cite{SER_QED}). The components $G^{(1)}_{\mu 4} = - \kappa T_{\mu 4}$ can therefore be used to eliminate degrees of freedom from the ten degrees of freedom  of $h_{\mu \nu}$ to be clarified in more detail below.

Instead of making use of  an extension of the Helmholtz decomposition of vector fields (see, e.g.,  Ref.\cite{SER_QED} for an elementary discussion and further references)  to tensor fields, we find it convenient and more transparent to analyze the corresponding constraints by considering the Fourier transform of Einstein's  equation (\ref{grav2})  in the weak-field limit. We therefore define  
\begin{equation} \label{grav7}
    T_{\mu \nu}(k) = \int d^4x e^{- i k\cdot x} T_{\mu \nu}(x) ~ ,   
\end{equation}
and 
\begin{equation} \label{grav8}
    h_{\mu \nu}(k) = \int d^4x e^{- i k\cdot x} h_{\mu \nu}(x) ~ .  
\end{equation}
Eqs.(\ref{grav2}) and (\ref{grav4}) therefore lead to 
\begin{equation} \label{grav9}
   \bigg ( \delta_{\mu}^{\alpha} \delta_{\nu}^{\beta} k^2 + \eta^{\alpha \beta} k_{\mu} k_{\nu} - \delta_{\nu}^{\beta} k_{\mu} k^{\alpha} - \delta_{\mu}^{\alpha} k_{\nu} k^{\beta} + \eta_{\mu \nu} \big ( k^{\alpha} k^{\beta} - \eta^{\alpha \beta} k^2 \big ) \bigg ) h_{\alpha \beta}(k) = 2 \kappa T_{\mu \nu}(k) ~ , 
\end{equation}
where $k^2 \equiv {\bf k}^2 - k_4^2$. The gauge-invariance of $G_{\mu\nu}^{(1)}$ under the transformation (\ref{grav6}) now allows us to introduce  four Coulomb-like  or radiation gauge conditions $\partial_ih_{i\mu}(x) =0$, i.e., 
\begin{equation} \label{grav10}
   k_i h_{i \mu}(k) = 0 ~ , 
\end{equation}
where Latin indices run from $1$ to $3$. With the gauge choice Eq.(\ref{grav10}), we find from Eq.(\ref{grav9}) the constraint equations
\begin{equation} \label{grav11}
    {\bf k}^2 h_{i4}(k) + k_i k_4 h_{ll}(k) = 2 \kappa T_{i4}(k) ~ ,
\end{equation}
as well as
\begin{equation} \label{grav12}
     {\bf k}^2 h_{44}(k) - k^2 h_{ll}(k) = \kappa T_{\mu}^{\mu}(k) ~ . 
\end{equation}
The transversality gauge-condition (\ref{grav10}), conservation of the energy-momentum tensor, i.e., $k_{\mu} T^{\mu \nu}(k)=0$, as well as Eq.(\ref{grav11}), then imply that ${\bf k}^2 k_4 h_{ll}(k) = 2 \kappa k_i T_{i4}(k) = 2 \kappa k_4 T_{44}(k)$. We  therefore obtain
\begin{equation} \label{grav13}
    {\bf k}^2 h_{ll}(k) = 2 \kappa T_{44}(k) ~ .
\end{equation}
By making use of Eq.(\ref{grav13}), we can therefore write Eq.(\ref{grav11}) in the following form where the transversality of $h_{i4}(k)$ is explicit:
\begin{equation}  \label{grav14}
    {\bf k}^2 h_{i4}(k) = 2 \kappa P_{ij} T_{j4}(k) ~ .
\end{equation}
Here $P_{ij}$ is given by 
\begin{equation}
\label{eq:polarization_sum}
P_{ij} \equiv \delta_{ij} - \hat{{ k}}_i\hat{{ k}}_j =\sum_{\lambda }{{\epsilon}}^*_i({\bf k} ;\lambda){{\epsilon}}_j({\bf k} ;\lambda)  \, \, ,
\end{equation}
expressed, for later purposes, in terms of polarization vectors ${\bm{\epsilon}}({\bf k} ;\lambda)$, with $\lambda = \pm$ for  complex-valued  circular polarization and  with $\lambda = 1,2$ for real-valued linear polarization,  obeying the transversality condition $\hat{{\bf k}} \cdot  {\bm{\epsilon}}({\bf k} ;\lambda) = 0$, and where we have defined the unit vector $\hat{{\bf k}} \equiv {\bf k}/|{\bf k}|$. The circular polarization degrees of freedom  obey the rule ${\bm{\epsilon}}(-{\bf k} ;\pm) = {\bm{\epsilon}}^*({\bf k} ;\pm)$.
In the case of linear polarization the   unit vectors ${\bm{\epsilon}}({\bf k} ;\lambda)$  are such that  ${\bm\epsilon}(-{\bf k} ;\lambda) = (-1)^{\lambda +1}{\bm\epsilon}({\bf k} ;\lambda)$ and 
\begin{equation}
\label{eq:circular_polarization}
{\bm{\epsilon}}({\bf k};\pm)= \frac{1}{\sqrt{2}}\big({\bm{\epsilon}}({\bf k} ;1) \pm i{\bm{\epsilon}}({\bf k} ;2)\big)\,\, . 
\end{equation}
 Conservation of the energy-momentum tensor allows us, furthermore, to rewrite the constraint equation (\ref{grav12}) in the form
\begin{equation}   \label{grav15}
   {\bf k}^2 h_{44}(k) = 2 \kappa \bigg ( P_{lm} T_{lm}(k) - \frac{1}{2} T_{\mu}^{\mu}(k) \bigg ) ~ .
\end{equation}
In the Newtonian limit, Eq.(\ref{grav15}) reduces to ${\bf k}^2 h_{44}(k)= \kappa T_{44}(k)= 8\pi G\rho(k)/c^2$ in terms of the mass density $\rho(k)$, i.e., $h_{44}(x) = -2\Phi (x)/c^2$, as it should,  where $\Phi (x)$ is the classical Newtonian gravitational potential such that $\nabla^2 \Phi (x)=4\pi G\rho(x)$. 

The components $h_{i4}(k)$ and $h_{44}(k)$ are therefore fixed by the constraint equations (\ref{grav14}) and  (\ref{grav15}). The spatial trace $h_{ll}(k)$ is, in addition, determined by these constraints according to Eq.(\ref{grav13}).  When transformed back to space-time coordinates these constraints will only involve spatial derivatives evaluated at equal time.    The constraint equation $G_{\mu 4}^{(1)} = - \kappa T_{\mu 4}$ has therefore been solved for. Here we remark that the Coulomb-like degrees of freedom $h_{i4}(x)$, $h_{44}(x)$, and $h_{ll}(x)$ are not retarded and, in fact, they are instantaneous in time.
\newline
\indent The dynamical Fourier transformed wave-equation for the components $h_{ij}(k)$ that follows from Eq.(\ref{grav9}) can conveniently now  be written in the following form: 
\begin{equation} \label{grav16}
    k^2 \bigg ( h_{ij}(k) - \frac{1}{2} P_{ij} h_{ll}(k) \bigg )  = 2\kappa P_{ij;lm}T_{lm}(k)~ .
\end{equation}  
Here we have defined the tensor projection operator $P_{ij;lm}$ by
\begin{equation}
\label{eq:grav_polarization_tensor}
 P_{ij;lm} \equiv \frac{1}{2} \bigg ( P_{il} P_{jm} + P_{jl} P_{im} - P_{ij} P_{lm} \bigg ) \,\, ,
\end{equation}
such that $P_{ij;lm} P_{lm;kn} =P_{ij;kn}$. 
Both sides of Eq.(\ref{grav16}) are now by construction transverse, symmetric, and traceless in the indices $ij$. The gauge-invariant combination  
\begin{equation}
\label{graviton_field}
   {\chi}_{ij}(k) \equiv h_{ij}(k) - \frac{1}{2} P_{ij} h_{ll}(k) ~ , 
\end{equation}
then contains two-degrees of freedom corresponding to the degrees of freedom of an {\it on-shell} spin-two massless field, i.e.,  a propagating gravitational field.
In terms of a formal inverse Fourier transform $\delta^T_{ij} \equiv (\delta_{ij}- \partial_i\partial_j/\nabla^2)$ of the projection operator  $P_{ij}$, as defined by Eq.(\ref{eq:polarization_sum}), the transverse, traceless, and gauge-invariant spatial components $h_{ij}^{TT}({\bf x},t)$ of the tensor   $h_{\mu\nu}({\bf x},t)$ can now be written in the well known  form (see, e.g., Ref.\cite{Misner_1973})
\begin{equation}
\label{TT-components}
h_{ij}^{TT}({\bf x},t) = \delta_{ij;kl}^{TT}h_{kl}({\bf x},t) ~ .
\end{equation}
Here we have defined the projection operator
\begin{equation}
\label{TT-components_b}
\delta_{ij;kl}^{TT} \equiv \delta^T_{ik}\delta^T_{jl } -\frac{1}{2}\delta^T_{ij}\delta^T_{kl} ~ ,
\end{equation}
which in Eq.(\ref{TT-components}), of course,  can be expressed in terms of explicit non-local space-integrals (see, e.g., Ref.\cite{SER_QED}).
It then follows that $h_{ij}^{TT}({\bf x},t)= {\chi}_{ij}({\bf x},t)$. Similarly, apart from a factor $2\kappa$,  the inverse Fourier transform of the right-hand side of Eq.(\ref{grav16}) corresponds to the transverse and traceless part $T_{ij}^{TT}({\bf x},t)= \delta_{ij;kl}^{TT}T_{kl}$ of the tensor $T_{ij}({\bf x},t)$. At this point one remarks that Eq.(\ref{grav16}) does not mathematically prescribe the space-time causal and/or retardation properties of ${\chi}_{ij}({\bf x},t)$ as, e.g., discussed by Rohrlich \cite{rohrlich_01}. In Section \ref{sec:causality} we will, however, verify that quantum mechanics leads to the correct causal and retarded behaviour of suitable expectation values of a second-quantized version of the field ${\chi}_{ij}({\bf x},t)$.
%
%
%
\vspace{0.2cm}
\bc{
\section{The Second-Quantized Gravitational Field}
\label{sec:multimode}
}\ec
\seqnoll

A second-quantized interaction picture version of the components ${\chi}_{ij}(x)$ in Eq.(\ref{graviton_field})  takes a conventional form, i.e., 
\begin{equation}  
\label{eq:chi}
  {\chi}_{ij}(x)  = \sum_{{\bf k}\lambda} \,  \sqrt{ \frac{2\hbar \kappa c^2}{V \omega_k} } \,  \, \bigg ( \, {\epsilon}_{ij}({\bf k};\lambda)a_{{\bf k} \lambda} \, e^{ik\cdot x} + {\epsilon}_{ij}^*({\bf k};\lambda)a^*_{{\bf k} \lambda} \, e^{-ik\cdot x} \, \bigg ) ~ ,
\end{equation}
with $\lambda = 1,2$ or $\lambda = \pm$ for gravitons with ${k\hskip -2pt\cdot\hskip -2pt x} = {\bf k}\hskip -2pt\cdot\hskip -2pt{\bf x} -\omega_kt$ using $k_4=\omega_k/c =|{\bf k}|$, analogous to the  second-quantized transverse electro-magnetic potential ${\bf A}_T(x)$ for photons in the Coulomb gauge (see, e.g., Ref.\cite{SER_QED}). The presence of the factor $2\kappa $ is similar to the appearance of the factor $1/2\varepsilon_0$ in the second-quantized electro-magnetic  field potential ${\bm A}_T(x)$  and also makes $ {\chi}_{ij}(x)$ dimensionless. An overall numerical factor is then determined from, e.g., a suitable canonical commutation relation (see, e.g., Refs.\cite{weinberg_1965, weinberg_1965b,Eriksson_1978}) or, as will be verified below, from consistency with the classical Einstein field equation in terms of expectation values.   It is assumed that the free-field Hamiltonian $H_0$ used in the definition of the interaction picture takes the form
\begin{equation}
\label{eq:free_hamiltonian}
H_0 = \sum_{{\bf k}\lambda}\hbar\omega_k(a^*_{{\bf k} \lambda} a_{{\bf k} \lambda}+1/2) 
 \,\, .
\end{equation}

The components of the two, in general complex-valued, polarization tensors ${\epsilon}_{ij}({\bf k};\lambda)$ in Eq.(\ref{eq:chi}) are normalized in such a way that
%
%
\begin{equation}
\label{eq:general_normalization}
     \text{Tr} \big [ {\epsilon}^*({\bf k};\lambda) {\epsilon}({\bf k};\lambda')  \big ] \equiv \sum_{k,l} {\epsilon}_{kl}^*({\bf k};\lambda) \epsilon_{lk}({\bf k};\lambda') = \delta_{\lambda \lambda'} ~ .
\end{equation}
An explicit representation in terms of the real-valued,  symmetric and transverse photon linear polarization unit vectors ${\bm \epsilon}({\bf k};\lambda)$, with $\lambda = 1,2$, in Section \ref{sec:Einstein} is 
\begin{equation} \label{eq:gravp_pol_1}
     {\epsilon}_{ij}({\bf k};1) =\frac{1}{\sqrt{2}}\big({ \epsilon}_i({\bf k};1){\epsilon}_j({\bf k};1)- { \epsilon}_i({\bf k};2){ \epsilon}_j({\bf k};2)\big) ~ ,
\end{equation}
and
\begin{equation}\label{eq:gravp_pol_2}
     {\epsilon}_{ij}({\bf k};2) =  \frac{1}{\sqrt{2}}\big({\epsilon}_i({\bf k};1){\epsilon}_j({\bf k};2)+{\epsilon}_i({\bf k};2){\epsilon}_j({\bf k};1)  \big)~ .
\end{equation}
With, e.g.,  ${\bf k}= (0,0,k)$ then the  matrix elements of ${\epsilon}_{ij}({\bf k};\lambda)$ are given by ${\epsilon}_{11}({\bf k};1)= -{\epsilon}_{22}({\bf k};1)= 1/\sqrt{2}$,  ${\epsilon}_{12}({\bf k};2)= {\epsilon}_{21}({\bf k};2)= 1/\sqrt{2}$, and ${\epsilon}_{i3}({\bf k};\lambda)=0$ for $i=1,2,3$ \cite{weinberg_1965}. By construction ${\epsilon}_{ij}({\bf k};\lambda)$ obey the rule ${\epsilon}_{ij}(-{\bf k};\lambda)=(-1)^{\lambda +1}{\epsilon}_{ij}({\bf k};\lambda)$.
In passing we notice that the tensor nature of a classical metric deviation ${\chi}_{ij}(x)$  has been reported in  a recent and remarkable precise test  in terms of  observed classical  gravitational waves from the binary black hole coalescence event GW170814 \cite{Abbott_5}.
In terms of the normalization Eq.(\ref{eq:general_normalization}), we remark that the tensor projection operator $P_{ij;lm}$ as defined by Eq.(\ref{eq:grav_polarization_tensor}) can be obtained as follows
\begin{equation}
\label{eq:grav_polarization_b}
 P_{ij;lm} = P_{ij;lm}(\hat{\bf k}) \equiv \sum_{\lambda=1,2} {\epsilon}_{ij}^*({\bf k};\lambda) {\epsilon}_{lm}({\bf k};\lambda) \,\, .
\end{equation}

If we express the quantum field ${\chi}_{ij}(x)$ in terms of complex circular polarization we can make use of  real-valued transverse  photon linear polarization unit vectors ${\bm \epsilon}({\bf k};\lambda)$ and Eq.(\ref{eq:circular_polarization}).
We can then write
\begin{equation}
\label{eq:grav_helicity}
{\epsilon}_{ij}({\bf k};\pm) \equiv {\epsilon}_{i}({\bf k};\pm){\epsilon}_{j}({\bf k};\pm) = \frac{1}{\sqrt{2}}\left({\epsilon}_{ij}({\bf k};1) \pm i{\epsilon}_{ij}({\bf k};2) \right)\,\, ,
\end{equation}
such that ${\epsilon}_{ij}^*({\bf k};\pm)={\epsilon}_{ij}(-{\bf k};\pm)$. The corresponding creation operators $a_{{\bf k} \pm}^*$ are given by
\begin{equation}
\label{eq:helicity_basis}
a_{{\bf k} \pm}^*= \frac{1}{\sqrt{2}}(a_{{\bf k}1}^* \pm i a_{{\bf k}2}^*)\,\, . 
\end{equation}
We  then observe  that, by construction, 
\begin{equation}
\label{eq:change_of_basis}
\sum_{\lambda =1,2} {\epsilon}_{ij}({\bf k};\lambda)a_{{\bf k}\lambda}^* = \sum_{\lambda =\pm} {\epsilon}_{ij}^*({\bf k};\lambda)a_{{\bf k}\lambda}^*\,\, ,
\end{equation}
 which simply states that the quantum field ${\chi}_{ij}(x)$ in Eq.(\ref{eq:chi}) does not depend on the actual realization of the choice of polarization degrees of freedom.
Under a rotation around the ${\bf  k}$-axis with an angle $\theta$ one  now readily finds that  $a_{{\bf k} \pm } \rightarrow \exp(\pm i2\theta)a_{{\bf k} \pm }$ due to the rotation properties of the polarization tensors ${\epsilon}_{ij}({\bf k};\lambda)$, for $\lambda = 1,2$. From this we conclude that single  graviton states $|{\bf k}\pm \rangle \equiv a^*_{{\bf k}\pm}|0 \rangle$ carry  intrinsic helicities $\pm 2\hbar$.  

In addition to the free field Hamiltonian $H_0$ in Eq.(\ref{eq:free_hamiltonian}), and analogous to the second-quantization of the electro-magnetic field in the Coulomb gauge,  we can now construct a momentum operator ${\bm P}$ and a helicity operator ${\bm \Sigma}$ according to
\begin{equation}
\label{eq:einstein_3}
{\bm P} \equiv \sum_{{\bf k}\lambda}\hbar{\bm k}a^*_{{\bf k} \lambda} a_{{\bf k} \lambda} 
 \,\, ,
\end{equation}
and
\begin{equation}
\label{eq:einstein_4}
{\bm \Sigma} \equiv   2 \hbar\sum_{{\bf k}} \hat{{\bf k}}\big(a^*_{{\bf k}+} a_{{\bf k} +} - a^*_{{\bf k}-} a_{{\bf k} -}\big)
 \,\, .
\end{equation}
The complete set of commuting operators $H_0$, ${\bm P}$, and ${\bm \Sigma}$ are now such that diagonal one-particle states $|{{\bf k}}{\pm}\rangle$  carry all the appropriate quantum numbers of a spin-two particle, i.e., a graviton. A complete set of physical Fock-states can then be generated in a  conventional manner. In addition to the intrinsic spin angular momentum ${\bm \Sigma}$, photon as well as graviton states can also carry conventional orbital  angular momentum ${\bm L}$, which for photons plays an important role in many current contexts (see, e.g., Ref.\cite{OAM_2003} and references cited therein), but will not be of concern in the present paper.

For our purposes, the Eqs.(\ref{eq:free_hamiltonian}),(\ref{eq:einstein_3}), and (\ref{eq:einstein_4}) are given by construction. Nevertheless, a   gauge-invariant energy-momentum tensor for the gravitational field ${\chi}_{ij}(x)$ can be found, as first discussed in various forms and in the context of classical gravitational waves by Einstein \cite{Einstein_1916},  and is defined by (also see, e.g., Refs.\cite{Misner_1973,Eriksson_1978, Landau_Lifshitz_1975,Ohanian_1963, Caroll_2004, Maggiore_2008,Balbusa_2016} for other considerations)
\begin{equation}
\label{eq:einstein_1}
t_{\mu\nu}(x) \equiv \frac{1}{4\kappa} \left( \partial_{\mu}{\chi}_{ij}(x)\partial_{\nu}{\chi}_{ij}(x) - \frac{1}{2}\eta_{\mu\nu} \partial^{\alpha}\chi_{ij}(x) \partial_{\alpha}\chi_{ij}(x)\right)
 \,\, .
\end{equation}
At the classical level, one procedure to motivate Eq.(\ref{eq:einstein_1}) is, e.g.,  to consider the Einstein tensor $G_{\mu\nu}$ to second-order in the deviation $h_{\mu\nu}$,  denoted by $G_{\mu\nu}^{(2)}$. An effective energy-momentum tensor for the gravitational degrees of freedom can then be defined by $-G_{\mu\nu}^{(2)}/\kappa$  (see, e.g., Ref.\cite{Caroll_2004}, Section 7.6,  and references cited therein), a procedure also used in various studies of low energy effective quantum gravity \cite{Donoghue_2015}.
 By a straightforward calculation in the gauge Eq.(\ref{grav10}), and by  considering the source-free limit $T_{\mu\nu}\rightarrow 0$, we then obtain a gauge-invariant energy-momentum tensor $t_{\mu\nu}^{(grav)}$ for the free gravitational field ${\chi}_{ij}(x)$, i.e.,
\begin{equation} 
\label{eq:einstein_1b}
t_{\mu\nu}^{(grav)}(x) \equiv -\frac{1}{2\kappa} \left({\chi}_{ij}(x)\partial_{\mu}\partial_{\nu}{\chi}_{ij}(x) - \frac{1}{2}\eta_{\mu\nu} \chi_{ij}(x) \partial^{\alpha}\partial_{\alpha}\chi_{ij}(x)\right)
 \,\, .
\end{equation}
With the normalization of second-quantized graviton field ${\chi}_{ij}(x)$ with two polarization degrees of freedom in Eq.(\ref{eq:chi}), an additional factor $1/2$ must then be included in Eq.(\ref{eq:einstein_1b}). When performing space-integration over the quantization volume $V$ one verifies that Eqs.(\ref{eq:einstein_1}) and (\ref{eq:einstein_1b}) then leads to the same result.
Conventional conservation of energy-momentum can now be expressed in terms  of expectation values of the effective and conserved  energy-momentum tensor $T_{\mu\nu}+\langle t_{\mu\nu}\rangle$.

The energy-momentum tensor $t_{\mu\nu}$ in Eq.(\ref{eq:einstein_1}) is,  by construction, now such  that the quantized gravitational field ${\chi}_{ij}$ in Eq.(\ref{eq:chi})  leads to the correct diagonal free-field Hamiltonian, i.e.,
\begin{equation}
\label{eq:einstein_2}
H_0 = \int_{V}d^3x t_{44}({\bf x},t) 
 \,\, ,
\end{equation}
analogous to  the case of the free second-quantized transverse electro-magnetic field ${\bm A}_T$. A momentum operator ${\bm P}$ can also be obtain from Eq.(\ref{eq:einstein_1}) in terms of ${\bm t}=(t^{41},t^{42},t^{43})$ according to 
\begin{equation}
{\bm P} = \frac{1}{c}\int_{V}d^3x \, {\bm t}({\bf x},t)  
 \,\, .
\end{equation}
The orbital angular momentum is then given by ${\bm L}= {\int d^3x {{\bm x}\times}{\bm t}}({\bf x},t)/c$.  The helicity operator  ${\bm \Sigma}$ can  also be expressed in terms of the of quantum field ${\chi}_{ij}$  by taking the properties of ${\chi}_{ij}$ under rotations  into account in a standard manner, but will not be of importance in the present context. The total angular momentum is then given ${\bm J}= {\bm L}+{\bm \Sigma}$ analogous  to the case of the second-quantized radiation field ${\bf A}_T(x)$.

We observe that in many classical expositions one neglect the second term in the energy-momentum Eq.(\ref{eq:einstein_1}) and  performs an averaging procedure over several characteristic wave-lengths of gravitational radiation (see, e.g., Refs.\cite{Flanagan_2005, Misner_1973, Ohanian_1963,Caroll_2004,Maggiore_2008}). We, however, follow a conventional second-quantized field theory procedure making use of Eq.(\ref{eq:einstein_1}) in order to verify, e.g., Eq.(\ref{eq:einstein_2}). Furthermore, and according to the Weinberg-Witten theorem \cite{WW_1980}, massless particles with helicity larger than one cannot carry a Lorentz-covariant energy-momentum tensor. As a theorem  this  appears to contradict the construction above. The theorem follows from a consideration of the rotational properties of  matrix elements like  $\langle {{-\bf k}}{\pm}|t_{\mu\nu}| {\bf k}\pm\rangle$ around the ${\bf k}$-axis. Due to the properties of the polarization tensor ${\epsilon}_{ij}({\bf k};\lambda)$  in Eq.(\ref{eq:chi}) it, however, follows from an explicit calculation of such on-shell matrix elements, making use of the definition  of $t_{\mu\nu}$ according to Eq.(\ref{eq:einstein_1}) and Eq.(\ref{eq:chi}) for the quantum field ${\chi}_{ij}(x)$, that such matrix elements actually vanish identically and in this sense the Weinberg-Witten theorem is therefore avoided. 

\vspace{0.5cm}
%
\bc{
\section{The Causality Issue}
\label{sec:causality}
}\ec
\seqnoll

The well-known interaction picture Hamiltonian for a classical current ${\bf j}({\bf x},t)$ interaction with the second-quantized and transverse electro-magnetic field ${\bf A}_T({\bf x},t)$ takes an analogous form for  the gravitational field $\chi_{ij}({\bf x},t)$ (see, e.g., Ref.\cite{weinberg_1965,Eriksson_1978}), i.e., 
\begin{gather} 
  H_I(t)  =  -\frac{1}{2} \int_V d^3x T_{ij}({\bf x},t)\chi_{ij}({\bf x},t)
 \nonumber \\
  = -\sum_{{\bf k} \lambda} \, \sqrt{ \frac{\hbar\kappa c^2}{2V\omega_k} } \,  \, \bigg ( {\epsilon}_{ij}({\bf k} ;\lambda) a_{{\bf k} \lambda} \, e^{- i \omega_k t} T_{ij}({\bf k},t) + {\epsilon}_{ij}^*({\bf k} ;\lambda)a^*_{{\bf k} \lambda} \, e^{i \omega_k t} T_{ij}^*({\bf k},t) \, \bigg ) ~ ,
\label{eq:multimode_grav_1}
\end{gather}
where we make use of the spatial Fourier transform
\begin{equation}
     T_{ij}({\bf k},t) \equiv \int_V d^3x ~ e^{i {\bf k} \cdot {\bf x}} \, T_{ij}({\bf x},t) ~ .
\end{equation}
The total Hamiltonian of the system will, in general, also involve instantaneous act-at-a-distance c-number valued Newtonian-like terms (see, e.g., Ref.\cite{Eriksson_1978}) which, however,  will not be of any concern in the present paper.

Due to the linear dependence of $a_{{\bf k} \lambda}$ and $a_{{\bf k} \lambda}^*$  in Eq.(\ref{eq:multimode_grav_1}), it is now straightforward  to exactly solve for the unitary quantum dynamics (see, e.g., Refs.\cite{Feynman_1951, Glauber_1951, Carruthers_65, Eriksson_1967} for early discussions). 
Indeed, with $\ket{\psi(t)}_I \equiv \exp(itH_0/\hbar)\ket{\psi(t)}$, where we for convenience make the choice $t_0 =0$ for the initial time, the state $\ket{\psi(t)}_I$ is such that, 
\begin{equation}
\label{eq:multimode_grav_2}
     i \hbar \, \frac{d\ket{\psi(t)}_I}{dt} = H_I(t) \ket{\psi(t)}_I ~ ,
\end{equation}
where for observables   
\begin{equation}
\label{eq:multimode_grav_3}
     {\cal O}_I(t) \equiv \exp(it H_0/\hbar) {\cal O}\exp(-it H_0/\hbar) ~ ,
\end{equation}
and where, in our case,  $H_0$ is given by Eq.(\ref{eq:einstein_2}).
The time-evolution   for $\ket{\psi(t)}_I$ is then given by
\begin{equation}
\label{eq:multimode_grav_4}
     \ket{\psi(t)}_I = \exp \left ( \frac{i}{\hbar} \phi(t) \right ) \, \exp \left ( - \frac{i}{\hbar} \int_{0}^t dt' \, H_I(t') \right ) \, \ket{\psi(0)} ~ ,     
\end{equation}
for any initial pure state $\ket{\psi(0)}$.
The $c$-number phase $\phi(t)$, which will be of no importance in our considerations, can be computed according to
\begin{equation}
\label{eq:multimode_grav_5}
    i \phi(t) = \frac{1}{2\hbar}  \int_{0}^t  dt'  \Big [ N(t') , H_I(t') \Big ]  = \frac{1}{2\hbar}  \int_{0}^t dt'\int_{0}^{t'} dt''  \Big [ H_I(t'') , H_I(t') \Big ]\,\,,
\end{equation}
with 
\begin{equation}\label{eq:multimode_grav_6}
      N(t) \equiv \int_0^t dt' \, H_I(t') ~ ,
\end{equation}
since $[ \, N(t') , H_I(t') \, ]$ now is a $c$-number. 

Apart from the  phase-factor $\phi(t)$, the time-evolution in the interaction picture can therefore be expressed in terms of  a conventional multi-mode displacement operator $U({\bm\alpha})$ as (see, e.g., Refs.\cite{Klauder_Skagerstam, Skagerstam_Klauder,Dodonov_Manko}) given by
\begin{equation}
\label{eq:multimode_grav_7}
 U({\bm\alpha}) \equiv  \exp \left ( - \frac{i}{\hbar} \int_{0}^t dt' \, H_I(t') \right )  = \prod_{{\bf k}\lambda} \, \exp \Big ( \alpha_{{\bf k}\lambda}(t) a^*_{{\bf k}\lambda} - {\alpha}^*_{{\bf k}\lambda}(t) a_{{\bf k}\lambda} \Big ) ~ ,
\end{equation}
where, in our case,
\begin{gather} 
   \alpha_{{\bf k}\lambda}(t) \equiv \frac{i}{\hbar} \, \sqrt{ \frac{\hbar\kappa c^2}{2V \omega_k} } \, \int_{0}^t dt' \, e^{i \omega_k t'}  {\epsilon}_{ij}^*({\bf k}; \lambda)T_{ij}^*({\bf k},t') \,\, .
\label{eq:multimode_grav_8}
\end{gather}
Since expectation values are independent of the picture used, i.e., 
\begin{equation}
   \langle {\cal O}(t) \rangle \equiv\,  _S\langle \psi(t)| {\cal O} |\psi(t) \rangle_S \, = \, _I\langle \psi(t)| {\cal O}_I(t) |\psi(t) \rangle_I  ~ ,
\end{equation}
 and by upon acting with the displacement operator $U({\bm \alpha})$ in Eq.(\ref{eq:multimode_grav_7})  on, e.g.,  an initial vacuum state, we can now evaluate the exact expectation value of second-quantized field $\chi_{ij}({\bf x},t)$ in a manner similar to the analysis for QED situation \cite{SER_QED}. We then obtain   the following $\hbar$-independent weak-field limit result
\begin{gather} 
\langle \chi_{ij}({\bf x},t) \rangle = 2\kappa c^2\int d^3x'
\sum_{{\bf k}} \, \frac{1}{V } \,  P_{ij;lm}(\hat{{\bf k}})e^{ i {\bf k} \cdot ({\bf x}-{\bf x}') } \, \int_0^t dt' \; \frac{ \sin \Big ( \omega_k (t-t') \Big ) }{\omega_k} T_{lm}({\bf x}',t')  \nonumber \\
 = 2\kappa c^2\int d^3x'\int_{0}^{t}dt' G_R({\bf x}-{\bf x}',t-t') T^{\text{TT}}_{ij}({\bf x}',t') \,\, ,
\label{eq:multimode_grav_9}
\end{gather}
in the large volume $V$ limit, with $\sum_{{\bf k}} =V\int d^3k/(2\pi)^3$, and making use of partial integrations. Here $G_R({\bf x},t)$ is the  retarded Green's function
\begin{equation} 
    G_R({\bf x},t) \equiv \int \frac{d^3k}{(2\pi)^3} e^{i{\bf k} \cdot {\bf x}} \frac{ \sin( \omega t ) }{\omega} \Theta(t) = \frac{\delta(t-|{\bf x}|/c)}{4\pi c^2 |{\bf x}|}  ~ ,
\label{eq:multimode_grav_10}
\end{equation}
for $t \geq 0$. Under time-reversal $t\rightarrow -t$ we remark that we can keep the expressions above with $t \rightarrow |t|$, i.e., we preserve  time-reversal invariance provided that the source $T_{ij}({\bf x},t)$ is also time-reversed.
In the derivation of Eq.(\ref{eq:multimode_grav_9}) we have made use of Eq.(\ref{eq:grav_polarization_b}) in the summation over the polarization degrees of freedom. 
A more explicitly form of Eq.(\ref{eq:multimode_grav_9}) is the well-known  causal and properly retarded expression
\begin{gather} 
\langle \chi_{ij}({\bf x},t) \rangle = \frac{4G}{c^4}\int d^3x' \frac{T^{\text{TT}}_{ij}({\bf x}',t- |{\bf x}-{\bf x}'|/c)}{|{\bf x}-{\bf x}'|} \,\, .
\label{eq:multimode_grav_11}
\end{gather}
The quantum mechanical expectation value $\langle \chi_{ij}({\bf x},t) \rangle$ in Eq.(\ref{eq:multimode_grav_11}) obeys the classical equation of motion 
\begin{gather} 
\square \langle \chi_{ij}({\bf x},t) \rangle = -2\kappa  T^{\text{TT}}_{ij}({\bf x},t) \, .
\label{eq:multimode_grav_12}
\end{gather}
Eq.(\ref{eq:multimode_grav_12}) is, of course, the Fourier transform of Einstein's classical dynamical Eq.(\ref{grav16}). If the initial state is changed from the vacuum state to an arbitrary pure state an additional contribution to Eq.(\ref{eq:multimode_grav_11}) appears corresponding to a homogeneous solution of the wave equation Eq.(\ref{grav16}) for $\chi_{ij}$. 
 Together with the constraint equations $G_{\mu 4}^{(1)} = - \kappa T_{\mu 4}$,   required to have the energy-momentum tensor $T_{\mu\nu}$ conserved,  the complete set of the classical and linearized Einstein's equations thus emerge from the quantum-mechanical  framework outlined above  analogous to the situation in QED \cite{SER_QED}.

The gauge-invariant Riemann-Christoffel tensor $R_{\lambda\mu\nu\kappa}({\bf x},t)$  in the weak field limit according to Eq.(\ref{Riemann_Chrisfoffel}) plays an important role in, e.g.,  the detection of gravitational waves (see, e.g., Refs.\cite{Weinberg_1972, Bertschinger_1996,Flanagan_2005,  Misner_1973, Landau_Lifshitz_1975, Ohanian_1963, Caroll_2004, Maggiore_2008, Creighton_2011}). Its causal and retardation properties in the far-field limit of a sufficiently localize energy-momentum tensor $T_{\mu\nu}({\bf x},t)$ are therefore of importance. In the case  of two test particles their spatial separation, i.e., the corresponding geodesic deviation,  is then determined by, e.g.,  the components $R_{l4 m 4}({\bf x},t)$ which we now express in terms of a space-time Fourier transform of Eq.(\ref{Riemann_Chrisfoffel}), i.e., 
\begin{gather} 
R^{(1)}_{l4 m 4}(k) = \frac{1}{2}\big(k^2_4h_{lm}(k)- k_lk_4h_{m4}(k)- k_mk_4h_{l4}(k)+ k_lk_mh_{44}(k)\big)\, .
\label{eq:multimode_grav_13}
\end{gather}
In terms of   $h_{lm}(k)= \chi_{lm}(k) +P_{lm}h_{nn}(k)/2$ we obtain
\begin{gather} 
R^{(1)}_{l4 m 4}(k) = \frac{k^2_4}{2}\chi_{lm}(k) + \frac{\kappa}{{\bf k}^2}\Big(\, \frac{k^2_4}{2}P_{lm}T_{44}(k) \nonumber \\
- k_4\big(\, k_mP_{lj}T_{j4}(k) + k_lP_{mj}T_{j4}(k) \big )+ k_lk_m\big(\, P_{ij}T_{ij}(k) - \frac{1}{2}T^{\mu}_{\mu}(k) \, \big) \, \Big)\, .
\label{eq:multimode_grav_14}
\end{gather}
where we have made use of the constraints Eqs.(\ref{grav13}), (\ref{grav14}), and(\ref{grav15}). When inverting the Fourier transform $R^{(1)}_{l4 m 4}(k)$ we make use of the fact that the integrals $\int d^3x T_{4m}({\bf x},t)$ are time-independent  for a sufficiently well localized energy-momentum tensor $T_{\mu\nu}({\bf x},t)$.  The far-field limit of  $R^{(1)}_{l 4 m 4}({\bf x},t)$ is then obtained from Eq.(\ref{eq:multimode_grav_14}) with the    properly causal and retarded gauge-invariant result 
\begin{gather} 
\langle R^{(1)}_{l4 m 4}({\bf x},t) \rangle = - \frac{1}{2c^2}\frac{\partial^2}{\partial t^2}\langle \chi_{lm}({\bf x},t) \rangle \, .
\label{eq:multimode_grav_15}
\end{gather}

For quantum fields in general, intrinsic quantum fluctuations are present \cite{Bohr_1950,Wightman_1964} and the related considerations for photons  applies also for quantum states of gravitons. Since  the Riemann-Christoffel tensor $R^{(1)}_{l 4 m 4}({\bf x},t)$ has now been promoted to an interaction-picture quantum-field,  we remark that intrinsic quantum fluctuations will be associated with expectation values like in Eq.(\ref{eq:multimode_grav_15}). As long as the equations of motion are linear  averaging procedures over a space-time resolution can be implemented as in the case of electro-magnetic interactions as discussed in, e.g., Ref. \cite{SER_QED}. Since gravity couples to an effective  conserved energy-momentum tensor non-linearities naturally emerge \cite{weinberg_1965,deser_75,Donoghue_2015} and the corresponding  averaging procedures require clarification which, however, goes beyond the scope of the present paper (in this context also see Ref.\cite{Unruh_2017}) .
\vspace{1.2cm}

\bc{
\section{\large Gravitational Radiation Processes}
\label{sec:grav_emission_proc}
}
\ec
\seqnoll

The emission of soft photons and/or gravitons \cite{weinberg_1965b,Eriksson_1978} can be investigated in terms of the formalism discussed above.
The classical expression for gravitational quadrupole emission from a given, sufficiently well localized, source can, in addition,   be obtained from the classical expression  Eq.(\ref{eq:multimode_grav_11}) in a standard manner (see, e.g., Refs.\cite{Weinberg_1972, Misner_1973, Landau_Lifshitz_1975, Ohanian_1963, Maggiore_2008, Creighton_2011}). 

We, however, notice the following. In view of diagonal form of the  free-field  Hamiltonian according to  Eqs.(\ref{eq:free_hamiltonian})  or (\ref{eq:einstein_2}), and the coherent state generated by the interaction Eq.(\ref{eq:multimode_grav_1}), it is now actually quite  easy  to find the time-dependence of the energy of the gravitons emitted. We first recall that conservation of the energy-momentum tensor $T_{\mu\nu}$ leads to the well-known identity 
\begin{gather} 
\label{eq:grp_1}
\int d^3x T_{ij}({\bf x},t) = \frac{1}{2c^2}\frac{\partial ^2}{\partial t^2}\int d^3xx_ix_j T_{44}({\bf x},t)\,\, ,
\end{gather}
for a sufficiently well-localized source $T_{\mu \nu}({\bf x},t)$.
By enforcing the dipole approximation 
\begin{gather} 
\label{eq:grp_2}
 T_{ij}({\bf k},t) \equiv \int d^3x e^{i{\bm k}\cdot{\bf x}}T_{ij}({\bf x},t) \approx \int d^3x T_{ij}({\bf x},t)\,\, ,
\end{gather}
and after a summation over polarizations using Eq.(\ref{eq:grav_polarization_b}), one finds, disregarding the divergent vacuum contribution in $H_0$ in Eq.(\ref{eq:free_hamiltonian}),  that
\begin{gather} 
E(t) \equiv  {_I}\langle\psi(t)| H_0 |\psi(t) \rangle_I = \sum_{{\bf k}}  \hbar \omega _k \left( \frac{1}{\hbar} \sqrt{\frac{\hbar\kappa c^2}{2V\omega_k}} \right)^2 
P_{ij;lm}(\hat{\bf k}) \nonumber \\\times\int_{t_0}^t dt'\int_{t_0}^t dt'' e^{i\omega_k(t'-t'')}
T_{ij}(t')T_{lm}(t'') ~ ,
\label{eq:grp_3}
\end{gather}
where used has been made of Eq.(\ref{eq:multimode_grav_8}).
Here we have defined 
\begin{gather}
T_{ij}(t) \equiv \int d^3x T_{ij}({\bf x},t) ~ ,
\label{eq:grp_4}
\end{gather}
which now can be expressed in terms of quadrupole moments according to Eq.(\ref{eq:grp_1}).
We now extend the  time interval to  $(-\infty,\infty)$, and define the Fourier transform $D_{kl}(\omega)$  of the moment $\int d^3xx_ix_kT_{44}({\bf x},t)/c^2$ in terms of the energy density $T_{44}({{\bf x}},t)=c^2\rho({\bf x},t)$ according to
\begin{gather} 
\label{eq:grp_5}
D_{kl}(\omega)  \equiv \int_{-\infty}^{\infty} \frac{dt}{2\pi} e^{i\omega t}D_{kl}(t) \,\, ,
\end{gather}
where
\begin{gather} 
\label{eq:grp_5b}
D_{kl}(t)  \equiv \int d^3 x x_kx_l\rho ({\bf x},t) \,\, .
\end{gather}
It then follows that
\begin{gather} 
\label{eq:grp_6}
E(t) = \frac{4\pi G}{5c^5}\int_{0}^{\infty }d\omega \omega ^6 Q^{*}_{kl}(\omega)Q_{kl}(\omega) \,\, .
\end{gather}
The angular average in Eq.(\ref{eq:grp_3}) has than been obtained by making use of  the result
\begin{gather} 
\label{eq:grp_7}
\int \frac{d\Omega}{4\pi} P_{ij;lm}(\hat{{\bf k}})D_{ij}(\omega)D^*_{lm}(\omega) =\frac{2}{5}Q^*_{ij}(\omega)Q_{ij}(\omega) \,\, ,
\end{gather}
expressed in terms of the gravitational quadrupole tensor $Q_{kl}(\omega)$ as defined by 
\begin{gather} \label{eq:C_7}
Q_{kl}(\omega) \equiv D_{kl}(\omega)- \frac{1}{3}\delta_{kl} D_{nn}(\omega) ~ .
\end{gather}
The total gravitational energy emitted  by the classical source according to Eq.(\ref{eq:grp_6}) agrees with well-known results (see, e.g., Ref.\cite{Weinberg_1972}, Section 10).

According to Eq.(\ref{eq:grp_3}) we now also observe  that
\begin{gather} 
E(t) = \frac{G}{5\pi c^5}{\int_{0}^{\infty}d\omega}{\int_{t_0}^t dt'}{\int_{t_0}^t dt''} \ddot{Q}_{ij}(t')\ddot{Q}_{ij}(t'') \frac{\partial ^2 }{\partial t' \partial t ''}e^{i\omega(t'-t'')}
 ~ ,
\label{eq:eformula_1}
\end{gather}
where the angular average has been obtained by making use of a similar expressions like  Eqs.(\ref{eq:grp_7}) for $D_{kl}(t)$ with
\begin{gather} \label{eq:C_7b}
Q_{kl}(t) \equiv D_{kl}(t)- \frac{1}{3}\delta_{kl} D_{nn}(t) ~ .
\end{gather}
If we now, e.g., imagine that the source is non-zero in the finite time-interval $[0,t]$ partial integrations can be carried out in Eq.(\ref{eq:eformula_1}),  and we  then obtain the famous Einstein expression \cite{Einstein_1916} quadrupole gravitational radiated power $P(t) \equiv d E(t) /dt$, i.e., 
\begin{gather} 
P(t) = \frac{G}{5\pi c^5}\int_{-\infty}^{\infty}d\omega \int_{t_0}^t dt' \dddot{Q}_{ij}(t)\dddot{Q}_{ij}(t') e^{i\omega(t'-t)} = \frac{G}{5 c^5}\dddot{Q}_{ij}(t)\dddot{Q}_{ij}(t)\, \ .
\label{eq:eformula_2}
\end{gather}
If this power will be observed at a large distance $R$ from the classical source, the time-parameter $t$ in Eq.(\ref{eq:eformula_2}) should, of course, be replaced by the retarded time $t- R/c$.

Furthermore, we can now also  make use of the interaction Eq.(\ref{eq:multimode_grav_1})  in order to evaluate the total decay rate  $\Gamma$ for the microscopic  transition $\ket{i} \rightarrow \ket{f}$, with  $\ket{i}=\ket{a_i}\otimes\ket{0}$ and $\ket{f} = \ket{a_f}\otimes \ket{{\bf k}\lambda}$, of  spontaneous emission of a graviton with the energy $\hbar \omega_k$ from an excited hydrogen atom in the initial atomic state $\ket{a_i}$ to the final atomic state $\ket{a_f}$ to first-order in time-dependent perturbation theory in a standard manner. We then make use of $T_{44}({\bf x},t)=m_ec^2\delta^{(3)}({\bf x}-{\bf x}(t))$ in Eq.(\ref{eq:grp_1}), where ${\bf x}(t)$ is the position of the electron at time $t$ in the interaction picture.  The relevant matrix element for  the spontaneous one-graviton transition from the hydrogen atomic state  $\ket{a_i}=\ket{nlm}=\ket{3dm}$ to the final atomic ground state $\ket{a_f} = \ket{1s}$ is then given by
\begin{gather} 
\label{eq:matrix_element}
\langle f|H_I(0) |i \rangle = \frac{m_e \omega_{if} ^2}{4c}\sqrt{\frac{16\pi G\hbar}{V\omega_k}}{\epsilon}_{lm}^*({\bf k};\lambda) \langle a_f| x_lx_m|a_i \rangle\,\, .
\end{gather}
Here $\hbar\omega_{if} \equiv E_i - E_f = \hbar \omega_k$ corresponds to energy conservation, and where, for the well localized quantum atomic states, we  make use of the dipole approximation ${\bf k}\cdot{\bf x}\approx 0$. 
By summing over all  polarization degrees of freedom and directions of the emitted graviton, making use of Eq.(\ref{eq:grav_polarization_b}), the decay rate $\Gamma$, for large time-scales $t$, with finite $|\omega_k -  \omega_{if}|t \gg 1$, can then be obtained in a straightforward manner with the result (see Appendix A)
\begin{gather} 
\label{eq:graviton_rate}
\Gamma \equiv \frac{2\pi}{\hbar^2}\int d^3k\frac{V}{(2\pi)^3} \sum_{\lambda}\delta(\omega_k - \omega_{if})|\langle i|H_I(0) |f \rangle|^2 = 
\frac{3^8}{2^{13}}\frac{Gm_e^2a_B^4\omega_{if}^5}{5\hbar c^5} \,\, ,
\end{gather}
in the large volume $V$ limit, independent of the energy-degeneracy quantum number $m$ and where $a_B= \hbar /\alpha m_ec$ is the Bohr radius for a reduced mass $\approx m_e$  of the atomic hydrogen two-body system.
This result agrees with Ref.\cite{Boughn_2006} and with the semi-classical result as reported in Ref.\cite{kiefer_2004},  but not with results as quoted in Refs.\cite{Weinberg_1972,scadron_1979}. If $n_{{\bf k}\lambda}$ gravitons are present in the initial state, the rate $\Gamma$ in Eq.(\ref{eq:graviton_rate}) should be multiplied by the factor $(1+n_{{\bf k}\lambda})$ corresponding to stimulated emission. Absorption graviton processes can be studied in a similar manner.
%
%
%
\vspace{0.2cm}

\bc{
\section{\large Final Remarks}
\label{sec:final_remarks}
}
\ec
\seqnoll

A quantum-mechanical  framework offers a platform to study causality and retardation issues in the classical theory of Maxwell \cite{SER_QED} as well as in the weak-field limit of Einstein's general theory of relativity. Second-quantization of physical degrees of freedom and current conservation for a classical source, leads to the well established classical theory for electro-magnetism in terms of expectation values of quantum fields. Inherent and unavoidable quantum fluctuations of these expectation values  will always  be present \cite{Bohr_1950,Wightman_1964,Unruh_2017}. As we have  now explicitly verified,  the same reasoning applies for  a second-quantized gravitational field around a flat metric in the presence   of a sufficiently weak  and conserved classical energy-momentum tensor. The weak-field limit of Einstein's general theory of relativity than naturally emerges   expressed also in terms of expectation values of quantum fields.

  The overwhelming experimental support for Maxwell's classical theory does not necessarily imply the existence of photons as quantum states and doubts on the existence of such quantum states are sometimes put forward (see, e.g., Ref.\cite{Lamb_95}). However, it is clear that the quantum-mechanical derivation of the classical theory necessarily  implies the existence of single particle quantum states corresponding to a photon. 
The recent experimental spectacular discovery of gravitational radiation \cite{Abbott_5,Abbott_1,Abbott_2,Abbott_3,Abbott_4,Abbott_6} is an additional verification of the correctness of Einstein's general theory of relativity. Outstanding as such observations are, they do not necessarily prove the existence of gravitons as physical quantum states. In view  of electro-magnetism and photons as quantum states, one would, however,   encounter a fundamental difficulty in understanding a possible physical non-existence of gravitons. This is so since the arguments we have presented concerning the emergence of the weak-field limit of Einstein's classical theory of gravity from quantum mechanics, follow the same line of reasoning as in the case of electro-magnetism \cite{SER_QED}. Therefore, as for the existence  of photons, they necessarily predict the existence of gravitons. Various arguments have, however,  been put forward that, due to the weak coupling of gravitons to matter, it may, nevertheless,  be impossible to detect single gravitons. 
In this context we observe that one may, in addition to the analysis of Dyson \cite{Dyson_2004}, also consider the possible  detectability of quantum states of gravitons in the form of, e.g.,  a Fock state $|n\rangle \equiv (a^{*}_{{\bf k} \lambda})^n|0\rangle/\sqrt{n!}$  for a macroscopically
 large number $n$. Such quantum states have, of course,   a trivial expectation value of the quantized gravitational field Eq.(\ref{eq:chi}) but non-zero intrinsic quantum fluctuations. Related discussions on the existence of gravitons can be found in, e.g., Refs.\cite{Boughn_2006,Boughn2_2006,Sokolowksi_2006}.

We have also  observed that various  classical $\hbar$-independent expressions for weak perturbations around a flat space-time easily and  naturally follow from the quantum-mechanical framework provided the classical source has a conserved energy-momentum tensor. In the weak-field limit  such expressions are supposed to be exact results. We now observe that the power spectrum of gravitons emitted from a classical source, with  a conserved energy-momentum tensor, is exactly obtained from first-order time-dependent perturbation theory as can be seen by inspection of Eq.(\ref{eq:grp_3}).  It may come as  a surprise that a first-order quantum-mechanical calculation can give an exact $\hbar$-independent answer. An explanation of this, as it seems,  remarkable fact can now be traced back to the factorization of the time-evolution operator in terms of a displacement operator for  quantum states in the interaction picture according to Eq.(\ref{eq:multimode_grav_4}) making use of Eq.(\ref{eq:multimode_grav_1}). The phase $\phi(t)$ then contains the non-perturbative effects of all higher-order corrections to the first-order result. 
As a matter of fact, similar features are  known to happen also in some other situations.  As is well-known, the famous differential cross-section for  Rutherford scattering can be obtained exactly by making use of the first-order Born approximation. All higher order corrections will then contribute with an overall phase for probability amplitudes which follows from the exact solution (see the excellent discussion in Ref.\cite{Gottfrid_66}). The classical Thomson cross-section for low-energy light scattering on a charged particle is also exactly obtained from a Born approximation due to the existence of an exact low-energy theorem in quantum electro-dynamics (see, e.g., the discussion in Section 8 of Ref.\cite{Itzykson_1972}).

As alluded to above, the weak coupling of gravitons to matter makes it perhaps of a more academic issue to discuss various features of quantum states for single graviton. Recently it has, however, been observed that  well-aligned pairs of photons, with the same helicity quantum numbers, can emulate the quantum-mechanical properties of spin-two gravitons \cite{Fernandez_2015}. In such a sense one may even consider entangled pairs of such emulated gravitons and, e.g.,  carry out EPR-correlation experiments in terms of the CHSH 
Bell-like inequalities \cite{Bell_65,Clauser_69} in a similar manner as in the famous Aspect et al. \cite{Aspect_99} realization of entangled pairs of photons. For photons the $\theta$-angle dependence under rotations around the ${\bf k}$-direction according to 
$a_{{\bf k} \pm } \rightarrow \exp(\pm i\theta)a_{{\bf k} \pm }$, using  the definition Eq.(\ref{eq:helicity_basis}), will then be replaced by $\theta \rightarrow 2\theta$ for emulated gravitons.  The corresponding quantum-mechanical CHSH Bell-like correlations to be measured for entangled pairs of emulated gravitons can then easily be worked out simply by the replacement $\theta \rightarrow 2\theta$ but will not be discussed in detail here \cite{Skagerstam_2017b}. Quantum optics for gravitons can therefore, at least, be contemplated in terms of emulated gravitons.

Finally, we speculate, that if the sources of the emulated  gravitons \cite{Fernandez_2015} can be described by a conserved energy-momentum tensor, the theoretical arguments presented in this exposition should apply. The characteristic coupling constant $\kappa$ could then, of course, take a completely different  value and thereby open up a possible window for experimental investigations of  classical as well quantum-mechanical features of emulated gravitons in a completely different context.
%
%
%
%
%
\newpage
%
%
%
%
%
\renewcommand{\thesection}{A}
\renewcommand{\section}[1]{\refstepcounter{sectionc}\vspace{0.0cm}
\setcounter{subsectionc}{0}\setcounter{subsubsectionc}{0}\noindent 
	{\bf\thesectionc. #1}}
\renewcommand{\theequation}{\thesection.\arabic{equation}}
\setcounter{section}{1}
\begin{center}
{\Large {\sc Appendix }}
\end{center}
\vspace{0.5cm}
%
%
%
%
\renewcommand{\thesection}{A}
\setcounter{section}{1}
\seqnoll
%
%
%
%
%
%
\begin{center}
   {\Large \bf \thesection. {Decay Processes}}
\end{center}

The  decay rate ${\Gamma}$ is defined  by
\begin{gather} \label{eq:D_1}
 {\Gamma}  \equiv \int d \Omega ~ \sum_{\lambda=1,2} ~ \frac{d \Gamma}{d \Omega} 
                         = G\frac{ m_e^2 \omega_{if}^5}{\hbar c^5} \, Q \,\,  ,
\end{gather}
where, for the one-graviton emission process of a hydrogen-like atom $\ket{3dm}\otimes \ket{0} \rightarrow \ket{1s}\otimes \ket{{\bf k}\lambda}$, we have defined
\begin{gather} \label{eq:D_2}
   Q  \equiv \int {\frac{d \Omega}{4\pi}}  \, D^*_{ij} D_{kl} ~ { \sum_{\lambda=1,2} \epsilon_{ij}({\bf k};\lambda)  \epsilon^*_{kl}({\bf k};\lambda) } ~ .
\end{gather}
The matrix elements $D_{ij}$ are then obtained from the transition matrix elements Eq.(\ref{eq:matrix_element}) in the main text, i.e., 
\begin{gather} \label{eq:D_3}
 D_{ij} \equiv \int d^3x \psi_{3dm}^{*}({\bf x})x_ix_j \psi_{1s}({\bf x})~ .
\end{gather}
In the case with $m=0$  the only non-zero components are then $D_{33}/2= -D_{11}=-D_{22}= a^2_B\sqrt{3^7/2^{15}}$ in terms of the Bohr radius $a_B$ as in Ref.\cite{Boughn_2006}. It can be verified that ${\Gamma}$ does not depend on the quantum number $m$ as must be the case.
The polarization sum in Eq.(\ref{eq:D_2}) is now analogous to Eq.(\ref{eq:grp_7}) in the main text where we now have defined the quadrupole matrix elements $Q_{kl}$ by
\begin{gather} \label{eq:D_5}
Q_{kl} \equiv D_{kl}- \frac{1}{3}\delta_{kl} D_{nn} ~ ,
\end{gather}
in terms of Eq.(\ref{eq:D_3}).
By combining Eqs.(\ref{eq:D_1})-(\ref{eq:D_5}) the result Eq.(\ref{eq:graviton_rate}) in the main text is  obtained. Even though such a decay is most likely impossible to observe, this analysis shows that we have properly normalized the graviton quantum field $\chi_{ij}$ in Eq.(\ref{eq:chi}) since the same analysis for a purely classical source leads to the well-known Einstein quadrupole formula as discussed in the main text.
%
%

%
%
%
%
%
%
\vspace{0.8cm}
%
%
%
%
%
%
\newpage
\begin{center}
   {\bf \large ACKNOWLEDGMENT}
\end{center}

 B.-S. S. expresses his gratitude to the late  J. D. Jackson for, in 2002, pointing out a sign error in an earlier publication (see Ref.\cite{jackson_02}) as well  as  to the CERN laboratory, and in particular the head of the CERN Scientific Information Service J.  Vigen, for providing hospitality during the final  preparation of the present work.
The research by B.-S. S.  was supported in part by Molde University College and NTNU. K.-E. E gratefully acknowledges hospitality at Chalmers University of Technology. The research of P. K. R. was supported  by  Molde University College. The authors are also grateful for the hospitality provided for at
Department of Space, Earth and Environment, Chalmers University of Technology, Sweden.
%
%
%
%
%
%
%
%
%
%
%
\begin{center}
   {\bf \large REFERENCES}
\end{center}
%
 
%
%
%
%
%
%
%
\end{document}